\documentclass[prb,aps,twocolumn,superscriptaddress,showpacs]{revtex4}
\usepackage{graphicx}
\usepackage{amssymb}
\begin{document}
\title{Influence of oxygen vacancy on the electronic structure of HfO$_2$ film}

\author{Deok-Yong Cho}
\author{Jae-Min Lee}
\author{S.-J. Oh}\email{sjoh@plaza.snu.ac.kr}
\affiliation{CSCMR \& School of Physics and Astronomy, Seoul
National University, Seoul 151-742, Korea}

\author{Hoyoung Jang}
\affiliation{eSSC \& Department of Physics, Pohang University of
Science and Technology, Pohang 790-784, Korea}
\author{J.-Y. Kim}
\affiliation{Pohang Accelerator Laboratory \& Pohang University of
Science and Technology, Pohang 790-784, Korea}

\author{J.-H. Park}
\affiliation{eSSC \& Department of Physics, Pohang University of
Science and Technology, Pohang 790-784, Korea} \affiliation{Pohang
Accelerator Laboratory \& Pohang University of Science and
Technology, Pohang 790-784, Korea}

\author{A. Tanaka}
\affiliation{Department of Quantum Matter, ADSM, Hiroshima
University, Higashi-Hiroshima 739-8526, Japan}

\begin{abstract}
We investigated the unoccupied part of the electronic structure of
the oxygen-deficient hafnium oxide (HfO$_{\sim1.8}$) using soft
x-ray absorption spectroscopy at O $K$ and Hf $N_3$ edges. Band-tail
states beneath the unoccupied Hf 5$d$ band are observed in the O
$K$-edge spectra; combined with ultraviolet photoemission spectrum,
this indicates the non-negligible occupation of Hf 5$d$ state.
However, Hf $N_3$-edge magnetic circular dichroism spectrum reveals
the absence of a long-range ferromagnetic spin order in the oxide.
Thus the small amount of $d$ electron gained by the vacancy
formation does not show inter-site correlation, contrary to a recent
report [M. Venkatesan {\it et al.}, Nature {\bf 430}, 630 (2004)].
\end{abstract}
 \maketitle

Oxygen vacancy in a large-bandgap oxide changes the electrical
characteristics by forming defect states inside the bandgap.
Depending on the position and density of the defect energy levels,
the chemical potential of the oxide - frequently identified with the
Fermi level (E$_{\rm F}$) - changes as in the case of doping with
other atomic species. Hence, the characterization of the defect
level is a prerequisite for understanding the electron transfer
across the oxide-semiconductor interface, because the energy barrier
height at the interface depends on E$_{\rm F}$ of the
oxide\cite{4yeo,4peacock2}. With regard to hafnium oxide, the energy
position of this defect level has been estimated
theoretically\cite{4Rob,4xiong,4gavartin} and
experimentally\cite{4takeuchi,4kerber} to be above the middle of the
bandgap, so E$_{\rm F}$ is shifted upward in the presence of oxygen
vacancies. Since E$_{\rm F}$ is determined from the charge
neutrality condition, it is largely influenced by the detailed
density of states (DOS) near the bandgap\cite{4peacock2}; in this
case, the intensity of the features near the conduction band (CB) is
expected to be higher than those near the valence band (VB).

Recent density functional theoretical (DFT)
studies\cite{4Rob,4xiong,4gavartin,4zheng,4kang,4cock} have revealed
that the defect state is the localized Hf $d$ state. The nonzero
occupation in Hf $d$ shell is very natural, because Hf atom no
longer donates electron to the adjacent vacant sites. If an
inter-site correlation existed between the $d$-electrons, even a
long-range correlation in the electronic configuration of each site
might emerge. Surprisingly, room-temperature long range
ferromagnetism (FM) was recently reported in an oxygen-deficient Hf
oxide film\cite{4ven,4coey,4hong}, although an experimental
counter-evidence shows the absence of FM\cite{4abraham}. Since the
measured magnetization is the macroscopic quantity, it might
significantly depend on the purity of the sample. Furthermore,
recent DFT studies have revealed that FM can be induced by a hafnium
vacancy rather than an oxygen vacancy, while the oxygen vacancy can
at the most yield paramagnetism\cite{4hfv1,4hfv2,4weng}. Thus, it is
important to investigate the possible long-range spin order in the
oxygen-deficient HfO$_{x<2}$ from a {\it microscopic} point of view,
in that the microscopic quantities such as the spin or angular
momentum drive the macroscopic properties, i.e., the magnetization.
Since these microscopic values are determined from the electronic
structure, the study on the local electron configurations of each
atomic species will provide the information that is not influenced
by a macroscopic deformation (domain interaction, dislocation, etc.)
or an intervention of impurities.

\begin{figure}
\begin{center}
\includegraphics[width=0.3\textwidth]{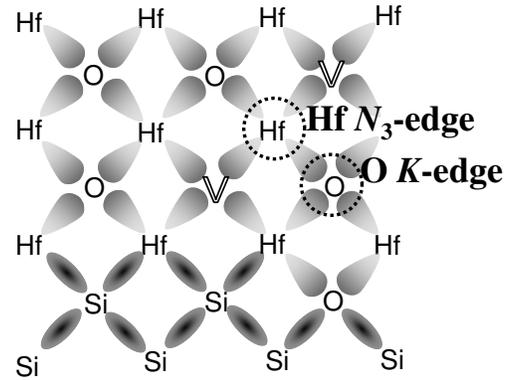}\end{center}
\caption[]{Schematic diagram of the atomic distribution of
HfO$_{x<2}$/Si. For simplicity, all the bonds are assumed to be
four-fold. Lobes indicate the inter-site wavefunction overlaps. The
shape and darkness in each lobe represent the electron density. The
dashed circles indicate the spatial extent to which each XAS
spectrum could detect the electronic structure. In O $K$-edge XAS,
it is impossible to observe the localized Hf $d$-electron state near
the oxygen vacancy (denoted as ``V"), if it were not for the
delocalization of the wavefunctions, in contrast to the Hf
$N_3$-edge XAS.}\label{fig:1}
\end{figure}

Therefore, in this paper, we focus on electron redistribution and
the possible long-range spin order in the presence of an oxygen
vacancy in Hf-oxide by studying its electronic structure. To
investigate the {\it unoccupied} electronic structure of Hf $d$
states, we performed x-ray absorption spectroscopy (XAS) study on
the HfO$_{x<2}$ film. Since XAS reflects a transition probability
from a spatially localized core-level to the unoccupied levels, it
is sensitive to the unoccupied states originating from the atoms
near the photon-absorbing atom. Further, owing to the ``locality"
that allows transitions irrespective of momentum transfer, it
reflects the momentum-averaged electronic structure that would be
equivalent to the partial density of states (PDOS) in the absence of
an electron correlation. We can choose the photon absorber by tuning
the photon energy range to any of the core-level energies of each
atomic species. Using the photon energy range $h\nu\sim530$ eV, we
can measure O $K$-edge ($1s\rightarrow2p$) absorption to investigate
the O 2$p$ PDOS. Since parts of O 2$p$ wavefunctions are hybridized
with Hf 5$d$ wavefunctions, O $K$-edge XAS also shows the unoccupied
Hf 5$d$ energy levels that are hybridized with the O wavefunctions.
This implies that an Hf $d$-electron state that stays near the
vacant site is hardly observed in O $K$-edge XAS. The situation is
described in Fig. \ref{fig:1}. The dashed circle near an oxygen atom
roughly indicates the probing area of O $K$-edge XAS. However, at Hf
$N_3$-edge ($4p_{3/2},h\nu\sim380$ eV), we can detect all of the Hf
5$d$ wavefunctions whether the $d$-electron state is localized near
the vacancy or hybridized with the surrounding oxygen atoms (see
Figure \ref{fig:1}). Although the obtained spectra become complex
due to the on-site core-hole effect, this measurement directly shows
the $d$-occupation at the Hf site.

The O $K$-edge and Hf $N_3$-edge XASs of the oxygen-deficient Hf
oxide were performed at the 2A EPU6 beamline of the Pohang Light
Source (PLS). In order to determine the existence of a long-range FM
spin order, we also performed magnetic circular dichroism (MCD)
measurements. The MCD spectrum is the difference between two XAS
spectra measured using external magnetic fields of mutually opposite
directions. Here, a circularly polarized photon should be used,
since the relative orientation of its chirality with respect to the
direction of the spin moment determines the preference of pumped
spins. To align the spin direction, the magnetic field exceeding the
coercivity field of the system should be applied. The
oxygen-deficient film of $\sim$20 {\AA} was fabricated onto a
HF-etched Si(100) substrate with a 3 Hz pulsed Nd:YAG laser at a
substrate temperature of 700 $^\circ$C. The base pressure of the
growth chamber was 1$\times10^{-8}$ Torr and no oxygen flow was
supplied during the film growth. The oxygen concentration deduced
from the intensity ratio of O 1$s$/Hf 4$f$ core-levels in the {\it
in situ} x-ray photoemission spectroscopy (XPS) spectra was
approximately 1.8 (i.e. 10\% oxygen deficiency). An approximate
estimation of the Hf valence leads to $\sim$Hf$^{3.6+}$, i.e., a
composite of 60\% of Hf$^{4+}$ and 40\% Hf$^{3+}$.

\begin{figure}
\begin{center}
\includegraphics[width=0.4\textwidth]{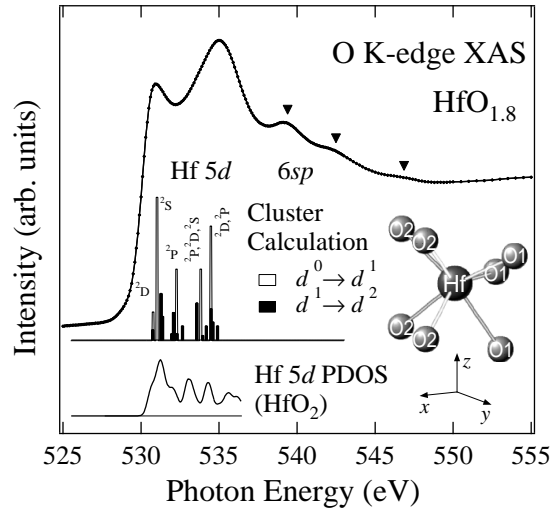}\end{center}
\caption[]{Unoccupied DOS of HfO$_{\sim1.8}$ probed by O $K$-edge
XAS with circularly polarized photons. The results of the atomic
single cluster model calculations [$d^0(d^1)\rightarrow d^1(d^2)$]
are also shown as empty (filled) bars. The cluster considered in
this study is shown in the inset. For comparison, the Hf 5$d$ PDOS
of HfO$_2$ taken from Ref. \cite{4peacock2} is shown at the bottom
of figure.}\label{fig:2}
\end{figure}

Figure \ref{fig:2} shows the O $K$-edge XAS spectrum with circularly
polarized photons for HfO$_{\sim1.8}$. The overall features of the
spectrum are similar to those of the previous experiments for
HfO$_2$\cite{4Toyoda,4lucov,4mhcho2} as well as the theoretical band
calculations for the unoccupied
states\cite{4foster,4muk,4xiong3,4cock}. The spectrum was analyzed
using a model cluster calculation with full atomic
multiplets\cite{4tanaka}, assuming the composite
[$d^0$(60\%)$\oplus$$d^1$(40\%)] in a perfect monoclinic crystal
structure. As already noted, XAS is relevant to the interactions
between the near atoms only; therefore, the single cluster model
with an Hf atom and its nearest oxygen atoms is sufficient to
simulate the XAS spectra. Though the Hf oxide film is in an
amorphous phase, the local bonds can be assumed to be similar to
those of the bulk monoclinic structure\cite{4cryst}. The crystal
field (CF) split energy ratios were calculated based on the bulk
crystal structure, and the overall CF splitting energy was used as a
variable scaling parameter to fit the experimental XAS spectra. The
one-electron energy level for Hf $5d^0\rightarrow 5d^1$ (empty bars
in Fig. \ref{fig:2}) is analyzed to be composed of five
non-degenerate states: $^2D(xy/x^2-y^2)$, $^2S(3z^2-r^2)$,
$^2P(xz/yz)$, and two mixed states in the increasing order of
energy. (Here the coordinate is taken as in Fig. \ref{fig:2}.) The
intensities of the five $d$ levels are multiplied by the Hf 5$d$$-$O
2$p$ hybridization strength using the Harrison's rule (exponent -7
for different interatomic distances)\cite{4Harrison} for the local
structure as shown in the figure. The hybridization strength was
found to exhibit an anisotropy; the in-planar $^2D$ state has much
weaker hybridization strength than the other levels ($\sim20$\% of
$^2S$ and $^2P$'s). The $5d^1\rightarrow 5d^2$ simulation (filled
bars) shows the similar results, except an overall broadening of
features due to the atomic multiplets in the $d^2$ final state
\cite{pointingout}. The spectra simulated with the overall CF energy
of approximately 4 eV, agree with the experimental spectrum and with
the calculated Hf 5$d$ PDOS\cite{4peacock2} shown in Fig.
\ref{fig:1}.

\begin{figure}
\begin{center}
\includegraphics[width=0.4\textwidth]{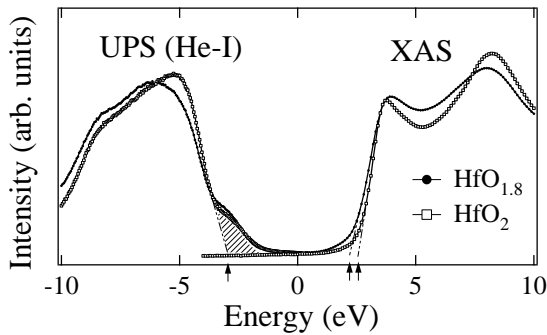}\end{center}
\caption[]{Combined (normalized) UPS and XAS spectra of
HfO$_{\sim1.8}$ and stoichiometric HfO$_2$, for DOS near the
bandgap. The UPS spectrum for HfO$_{1.8}$ is taken from Ref.
\cite{4cho3}. Here, the position of the chemical potentials (zero
energy in the figure) for both UPS and XAS spectra were arbitrarily
chosen to be approximately in the middle of the bandgap, preserving
the bandgap of $bulk$ HfO$_2$($\sim$5.7eV). Dotted lines show the
extrapolations to obtain the approximate energy positions (arrows)
of the CB and VB edges, and the shaded area indicates the
contributions of the Si substrate or the interface
state.}\label{fig:3}
\end{figure}
\begin{figure}
\begin{center}
\includegraphics[width=0.4\textwidth]{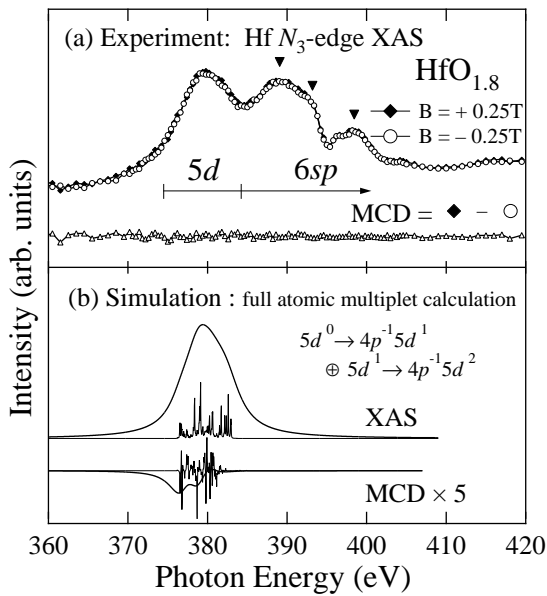}\end{center}
\caption[]{(a) Hf $N_3$-edge XAS with circularly polarized photons.
The difference spectrum or MCD has no feature, indicating no
long-range FM spin-order. The results of the atomic cluster model
calculations for the composite of $d^0$(60\%)$\oplus d^1$(40\%) are
also shown in (b).}\label{fig:4}
\end{figure}

A slight difference in the XAS spectra of the oxygen-deficient film
from the stoichiometric film can be seen in the features near the
bandgap as shown in Fig. \ref{fig:3}. For comparison, we appended
the XAS spectrum of the stoichiometric HfO$_2$ film measured under
the same conditions. Both spectra were normalized approximately
preserving the total absorption intensity. The spectrum of
oxygen-deficient film appears broader than that of the
stoichiometric one. We checked the broadness for many measurement
points on each sample and found the consistency for each sample.
Thus, the broadness can be regarded as an intrinsic property of the
oxygen-deficient film. Though no prominent gap state is observed in
the XAS spectrum of the oxygen-deficient film, the CB edge extends
toward the bandgap as compared to the stoichiometric one. This can
be interpreted as the partial delocalization of the defect state;
that is, the defect state might be rather dispersed so that it can
also be detected at the oxygen site. Thus, the role of the oxygen
vacancy can be regarded as donating electron back to the Hf site and
consequently to the O site through the Hf$-$O hybridization. In
particular, when the vacancy is at the HfO$_2$/Si interface, it is
easily substituted by the Si atom to form an O$-$Hf$-$Si bonding.
This could also slightly enhance the Hf $d$ occupation, although its
amount is reduced due to the electron transfer to the Si
site\cite{4Joo}. However, the CB tail cannot be attributed to the
formation of a direct O$-$Si bonding. This is because the bandgap of
SiO$_2$ ($\sim$9 eV) is much larger than of HfO$_2$ ($\sim$5.7 eV);
therefore, no structure can exist near the bandgap. We can expect
that the bandgaps of any types of the their intermediate state, i.e.
the silicate (Hf$_x$Si$_y$O$_4$) will be between these two
values\cite{4kato,4kami}, resulting in the absence of states inside
the HfO$_2$ bandgap. Therefore, it is evident that the lack of
oxygen atom in the oxide or at the interface is responsible for the
occurrence of the CB tail state\cite{4xiong3}.

If degree of amorphousness in the Hf-oxide is somehow enhanced, this
could also extend the CB edge due to the broadening of the
features\cite{4anderson}. In this case, the tailing should occur in
both the VB and CB edges. In the following paragraph, however, we
will show that the tailing does not occur in the VB edge; this
manifest that the CB tail is truly caused by the physics of defects,
not by the uncertainty in the lattice dynamics\cite{4anderson}.

To verify the existence of the band-tails in the {\it occupied} DOS,
we also performed ultraviolet photoemission spectroscopy (UPS) using
He$-$I ($h\nu=21.2$eV) light source for both the films. (The UPS
spectrum for HfO$_{\sim1.8}$ is taken from Ref. \cite{4cho3}.) The
UPS spectra are also normalized with respect to the total intensity.
Here, the energy zero is arbitrarily chosen to approximately sustain
the energy difference between the lowest XAS edge ($\sim$+2.5 eV)
and the highest UPS main feature ($\lesssim$-3 eV) to be the known
bandgap of HfO$_2$($\sim$5.7 eV; Ref. \cite{4bandgap}). The UPS
spectrum of HfO$_{\sim1.8}$ film is shifted by +0.3 eV in order to
compensate for the E$_{\rm F}$ shift in the oxide, which was
obtained from the cutoff kinetic energies of the secondary electrons
in the spectra (not shown here)\cite{4cho3}; this E$_{\rm F}$ shift
induces the higher binding energy (BE) shift in the photoemission
spectra, because the BE is, by definition, the energy difference
between the Fermi level and the core level, while XAS is independent
of the E$_{\rm F}$ position because the energy difference between
the initial core level and the final conduction level is independent
of the Fermi level. The large and broad features around the relative
energy of -8 $\sim$ -4 eV correspond to the main O $2p$ occupied
state, while the tail toward the bandgap (shaded area at around -3
eV in Fig. \ref{fig:3}) probably originates from the underlying Si
wafer\cite{4sayan}, because this VB tail is absent in the UPS
spectra of thicker ($>100$ {\AA}) HfO$_2$ film (not shown here). A
slight increase in the tail in the oxygen deficient film might be a
signature of the metallic Hf silicide (Hf$-$Si bond) at the
interface\cite{4cho1}. However, the VB edge of the oxygen-deficient
film does not extend inside the bandgap as compared with that of the
stoichiometric sample. Though it is difficult to unambiguously
analyze these band-tail states in both the occupied and unoccupied
DOSs, we can observe that only the unoccupied DOS near the bandgap
tends to extend inside the bandgap in the presence of oxygen
vacancy\cite{4cock,4foster}. This result should be correlated with
the lifting of Fermi level because the virtual transition to CB
becomes easier.

Up to this point, we have shown experimental evidence that the
oxygen vacancy lowers the CB edge and increases the effective Fermi
level. The detailed mechanism explaining how the vacancy formation
lowers the CB edge could be revealed by a further consideration of
the atomic rearrangements or the relaxation following the vacancy
formation. For example, the disorder in the positions of the vacant
sites can play a role in broadening the CB DOS and consequently in
reducing the bandgap. Further, the energy position of the defect
level is found to depend on the atomic arrangements after the
vacancy formation\cite{4xiong}. The reduction in the mean Hf$-$O
bond length ($\sim$-0.1 {\AA}) and in the number of bonds, in the
presence of oxygen vacancy\cite{4cho3}, will influence the detailed
balance in the charge dynamics between the two atomic species. The
CB lowering can also be correlated with the broadening of features
influenced by the valence configurations (the number of electrons)
of the remaining atoms, as shown in the simulation results in Fig.
\ref{fig:2}. Further investigation is required to completely
understand the mechanism.

Now we check if there be a long-range FM spin order. The Hf
$N_3$-edge XAS spectra obtained with circularly polarized photons
are shown in Fig. \ref{fig:4}(a). The external magnetic field was
maintained perpendicular to the sample plane with $H= \pm$2500 Oe,
which is much larger than the reported ferromagnetic coercivity
field of the HfO$_2$ film\cite{4ven}. The difference spectrum, i.e.,
the MCD spectrum exhibits no clear feature and remains zero within
the experimental uncertainty. This indicates that there is no Hf
$d$-electron-related FM spin order. One would argue that the density
of the oxygen vacancy is too small to detect FM; however, the value
of 10\% is larger than the dopant density in conventional diluted
magnetic semiconductors (DMS)\cite{4coey2}.

To analyze the spectral features, we appended in Fig. \ref{fig:4}(b)
the simulation result obtained by the full atomic multiplet
calculation for Hf $N_3$-edge XAS ($4p_{3/2}\rightarrow5d$) with the
parameters similar to that of O $K$-edge XAS, except the $4p-5d$
Coulomb interactions\cite{4Cowan}. As was in the O $K$-edge XAS
simulation, the simulated spectra are obtained by the sum of two
transitions: $5d^0\rightarrow4p^55d^1$ and
$5d^1\rightarrow4p^55d^2$. The features of the latter are shifted by
$U-F^0_{pd}=-2$ eV\cite{4Cowan} with respect to those of the former.
The result of the calculation is in an excellent agreement with the
experimental spectra; therefore, it provides definite peak
assignments; the Hf $4p_{3/2}{\rightarrow}5d$ transition is the
largest feature, while the other features in the higher photon
energies are attributed to the Hf $4p_{3/2}{\rightarrow}6sp$
transition. Contrary to $2p\rightarrow3d$ transition (as in the case
of transition metal ions), the feature of Hf $4p\rightarrow5d$
transition is not much stronger than $4p\rightarrow6sp$, because of
its lower photo-absorption cross-sections. The features of Hf $6sp$
for both the edges are denoted by $\blacktriangledown$ in Figs.
\ref{fig:2} and \ref{fig:4}. The peak positions almost coincide,
confirming our peak assignments. The simulated MCD signal for Hf
5$d$ [Fig. \ref{fig:4}(b)] is small due to the small anisotropy in
the unoccupied electron states ($d^{8-9}$). Although the sensitivity
of the MCD measurement on magnetism is low, no correlation was
observed between the experimental and simulated MCD spectra. This
confirms that the small modulation in the experimental spectrum is
caused by the noise, not the possible long-range magnetism.

Therefore, we can conclude that there is no preference in spin
selection in the electron occupation to either Hf $5d$ or $6sp$;
this evidently proves ``no ferromagnetic spin-order" in the case of
oxygen vacancy. However, the electron occupation number at an Hf
site should be nonzero; otherwise, the electronic configurations of
the next nearest O sites, which was shown in O $K$-edge XAS spectra
(Fig. \ref{fig:3}), should never be influenced by the presence of
oxygen vacancy. Thus, the absence of FM spin-order can be explained
by the absence of long-range correlation between the $d$-electrons
in the neighboring Hf sites, rather than the absence of $d$-electron
itself.

We conclude our arguments with two remarks: the oxygen vacancy in Hf
oxide slightly enhances the CB DOS. However, it does not involve the
long-range ferromagnetic spin order in the oxide because of the
absence of inter-site electron correlation.

This study is supported by the Korean Science and Engineering
Foundation through the Center for Strongly Correlated Materials
Research at the Seoul National University, and the experiments at
PLS were supported in part by MOST and POSTECH.


\begin{thebibliography}{99}
\bibitem{pointingout}
Here the effect of charges generated by the oxygen vacancy is
neglected, so that the overall CF splitting energy is assumed to be
independent of the number of $d$-electrons.
\end{thebibliography}
\end{document}